\documentclass[10pt,a4paper,twoside]{article}
\usepackage{amsmath}
\usepackage{amssymb}
\usepackage{cite}       

            {\hfil\hfill$\square$\endtrivlist}
\newtheorem{theorem}{Theorem}

\newcounter{note}
\newcommand\note{\refstepcounter{note}\trivlist
	\item[\hskip\labelsep{\textbf{\thenote. }}]}

\begin{document}
%
%
\pagestyle{myheadings}
\thispagestyle{plain}
\markboth{\hfill S. B. Edgar and A. H\"oglund\hfill}{\small The Lanczos potential for
Weyl-Candidate tensors exists only in four dimensions}

\begin{center}
	\huge\bfseries
	The Lanczos potential for Weyl-candidate tensors exists only
	in four dimensions
\end{center}

\begin{center}
	{\large
	S. B. Edgar\footnote{email: bredg@mai.liu.se}
	and A. H\"oglund\footnote{email: anhog@mai.liu.se}
	}\\[1mm]
	Matematiska institutionen, Link\"opings universitet,\\
	SE-581 83 Link\"oping, Sweden.
\end{center}

\begin{abstract}
We prove that a Lanczos potential $L_{abc}$ for the Weyl candidate
tensor $W_{abcd}$ does not generally exist for dimensions higher than
four. The technique is simply to assume the existence of such a
potential in dimension $n$, and then check the integrability
conditions for the assumed system of differential equations; if the
integrability conditions yield another non-trivial differential system
for $L_{abc}$ and $W_{abcd}$, then this system's integrability
conditions should be checked; and so on. When we find a non-trivial
condition {\it involving only $W_{abcd}$ and its derivatives}, then
clearly Weyl candidate tensors failing to satisfy that condition
cannot be written in terms of a Lanczos potential $L_{abc}$.
\end{abstract}

\section{Introduction}

{\it The mere presence of an unadulterated Riemannian geometry of specifically
4-dimensions brings into existence a tensor of third order of 16 components,
which bridges the gap between the second-order tensor of the line element and
the fourth order tensor of the Riemannian curvature.}
\ C. Lanczos (1962) Rev. Mod. Phys. {\bf 34}, 379.

Even though there is a deeper understanding today of the 3-tensor
potential for the Weyl tensor proposed by Lanczos~\cite{lanczos62},
there exists still a question mark over whether its existence is
unique to 4-dimensions, as seems to be claimed by Lanczos.

In this paper we demonstrate that it is not possible, in general, to
obtain a Lanczos potential for all Weyl candidates in spaces of
dimension $n$, where $n>4$; this does not prove unambiguously the
truth of Lanczos' claim [for the \emph{Weyl curvature tensor}], but
certainly emphasises that four dimensional spaces play a very special
role regarding the existence of Lanczos potentials.

Lanczos proposed that, in four dimensions, the Weyl tensor
$C_{abcd}$  could be given locally  in terms of a potential $L_{abc}$ as
\begin{equation}\label{WLorig}
	C^{ab}{}_{cd}
	=
	L^{ab}{}_{[c;d]}
	+L_{cd}{}^{[a;b]}
	-{}^*\!L^*{}^{ab}{}_{[c;d]}
	-{}^*\!L^*_{cd}{}^{[a;b]}
\end{equation}
where
\begin{subequations}\label{Lsym}
\begin{gather}
	L_{abc} = L_{[ab]c} \label{Lasym} \\
	L_{[abc]} = 0 \label{Lcycl}
\end{gather}
\end{subequations}
and $^*$ denotes the usual Hodge dual.  \label{fourdimproof} Bampi and
Caviglia~\cite{bampi83} have shown that, although his proof was
flawed, Lanczos' result is still valid. Moreover, Bampi and
Caviglia~\cite{bampi83} have shown that a Weyl candidate, i.e. \emph{
any} 4-tensor $W_{abcd} $ having the index symmetries
\begin{subequations}\label{Wsym}
\begin{gather}
	W_{abcd} = W_{[ab]cd} = W_{ab[cd]}, \label{Wasym} \\
	W_{abcd} = W_{cdab} \label{Wpsym} \\
	W_{a[bcd]} = 0 \label{Wcycl} \\
	W^a{}_{bad} =  0 \label{Wtrace}
\end{gather}
\end{subequations}
can be given locally in such a form, in four
dimensions. Illge~\cite{illge88} has given equivalent existence
results (as part of a larger investigation), using spinors, in
4-dimensional spacetimes with Lorentz signature.  Andersson and
Edgar~\cite{andersson00} has also given an existence proof in spinors.
They have also translated that proof into tensors in four dimensions; to
modify this proof to other dimensions seems impossible since some
identities valid only in four dimensions were needed in that proof.

Although the form given in \eqref{WLorig} clearly cannot be
generalised directly to other dimensions, yet by writing
\eqref{WLorig} in an equivalent form, without Hodge duals, it is
straightforward to generalise this equivalent form of \eqref{WLorig}
to $n>4$ dimensions,
\begin{equation}\label{WL}
\begin{split}
	W^{ab}{}_{cd}
	& =
	\  2  L^{ab}{}_{[c;d]}
	+2L_{cd}{}^{[a;b]}
	\\ &
	-\frac{4}{n-2}\delta^{[a}_{[c}\left(
		L^{b]}{}^i{}_{d];i}
		-L^{b]i}{}_{|i|;d]}
		+L_{d]i}{}^{b];i}
		-L_{d]i}{}^{|i|;b]}\right)
	\\ &
	+\frac{8}{(n-2)(n-1)} \delta^{a}_{[c}\delta_{d]}^{b}L^{ij}{}_{i;j}
\end{split}
\end{equation}
where $W_{abcd} $ and $L_{abc} $ satisfy \eqref{Wsym} and \eqref{Lsym}
respectively, (Bampi and Caviglia~\cite{bampi83}).

This can be further simplified to
\begin{equation}\label{WLsimp}
	W^{ab}{}_{cd}
	=
	2  L^{ab}{}_{[c;d]}
	+2L_{cd}{}^{[a;b]}
	-\frac{4}{(n-2)}\delta^{[a}_{[c}\left(
		L^{b]}{}^i{}_{d];i}
		+L_{d]i}{}^{b];i} \right)
\end{equation}
by the gauge choice
\begin{equation}
	L_{ab}{}^b = 0
\end{equation}
which is called the Lanczos algebraic gauge.  To see that this really
is a gauge choice , rather than a restriction, we can take any
$W_{abcd}$ and any $L_{abc}$ that satisfies \eqref{WL}, then
$L'{}_{abc} = L_{abc} + \Phi_{[a}g_{b]c}$ also satisfy \eqref{WL}
for all $\Phi_a$.  This is easily shown by inserting $L'{}_{abc}$ into
\eqref{WL}.  The particular choice $\Phi_a = -\frac{2}{n-1}L_{ab}{}^b$
gives the Lanczos algebraic gauge.  In the rest of this paper this
gauge choice as assumed.

From the existence proofs in four dimensions it follows that there is
a further gauge choice
\begin{equation}
	L_{abc}{}^{;c} = \xi_{ab}
\end{equation}
where $\xi_{ab}$ is any antisymmetric tensor.  When $\xi_{ab}=0$ this
choice is called the Lanczos differential gauge.  Whether this gauge
freedom remains in higher dimensions or not is not known.  Therefore
we have not assumed anything about this gauge in this paper.

Lanczos gave no direct support for his claim regarding the privileged
role of four dimensions, although there is some indirect evidence (see
Comment~\ref{note:nosurp} in section~\ref{sec:notes}) that Lanczos
potentials may not exist in higher dimensions. So we now consider the
question whether {\it every} 4-tensor $W_{abcd}$ having the index
symmetries \eqref{Wsym} can be given locally in the form \eqref{WL} in
terms of some $L_{abc}$ with the properties \eqref{Lsym}, {\it for all
spaces of dimension $n$, where $n>4$}.

The case when $n<4$ is not considered.  Any tensor $W_{abcd}$ with the
symmetries \eqref{Wasym} and \eqref{Wtrace} is identically zero in
those cases.  The right hand side of \eqref{WL} also has those
symmetries whenever $L_{abc}$ has the symmetry \eqref{Lasym}.
Therefore, any tensor $L_{abc}$ which is antisymmetric over its first
two indices satisfies \eqref{WL} identically when $n<4$.

\section{Lanczos potentials for Weyl candidates in Flat Space}

We start our considerations in flat space.  This investigation is important
in its own right, but it also enables us to get a pattern for how to
proceed in curved space.


\subsection{Six dimensions and higher}

By differentiating \eqref{WLsimp} twice we can
obtain,
\begin{equation}\label{WLd2c0flat}
	W^{[ab}{}_{[cd;e]}{}^{f]}
	=
	- \frac{4}{n-2}\delta^{[a}_{[c}
	\left(
		L^{b|i|}{}_{d;|i|e]}{}^{f]}
		+L_{d}{}^{|i|b}{}_{;|i|e]}{}^{f]}
	\right)
\end{equation}
Contracting this once gives
\begin{equation}\label{WLd2c1flat}
\begin{split}
	W^{[ab}{}_{[cd;i]}{}^{i]}
	& =
	-\frac{4(n-4)}{9(n-2)}
	\Bigl(
		L^{[a|i|}{}_{[c;|i|d]}{}^{b]}
		+L_{[c}{}^{i[a}{}_{;|i|d]}{}^{b]}
	\Bigr)
	\\ &
	-\frac{4}{9(n-2)}\delta^{[a}_{[c}
	\Bigl(
		L^{b]i}{}_{d];ij}{}^j
		+L_{d]}{}^{|i|b]}{}_{;ij}{}^j
	\\ &
		-L^{|ji|}{}_{d];ij}{}^{b]}
		-L_{d]}{}^{|ij|}{}_{;ij}{}^{b]}
		-L^{b]i}{}_{|j;i|d]}{}^j
		-L_{|j}{}^{|i|b]}{}_{;i|d]}{}^j
	\Bigr)
\end{split}
\end{equation}
and once more gives
\begin{equation}\label{WLd2c2flat}
\begin{split}
	W^{ai}{}_{cj;i}{}^j
	=
	\frac{(n-3)}{(n-2)}
	\Bigl( &
		L^{ai}{}_{c;ij}{}^j
		+L_{c}{}^{ia}{}_{;ij}{}^j
		-L^{ji}{}_{c;ij}{}^{a}
		-L_{c}{}^{ij}{}_{;ij}{}^{a}
	\\&
		-L^{ai}{}_{j;ic}{}^j
		-L_{j}{}^{ia}{}_{;ic}{}^j
	\Bigr)
\end{split}
\end{equation}
This can be substituted into \eqref{WLd2c1flat} to give
\begin{equation}\label{WLd2c1sflat}
\begin{split}
	W^{[ab}{}_{[cd;i]}{}^{i]}
	& =
	-\frac{4(n-4)}{9(n-2)}
	\left(
		L^{[a|i|}{}_{[c;|i|d]}{}^{b]}
		+L_{[c}{}^{i[a}{}_{;|i|d]}{}^{b]}
	\right)
	\\ &
	-\frac{4}{9(n-3)}\delta^{[a}_{[c} W^{b]i}{}_{d]j;i}{}^j
\end{split}
\end{equation}
which in turn can be substituted into \eqref{WLd2c0flat} to give
\begin{equation}\label{nocyclicrestrict}
\begin{split}
	W^{[ab}{}_{[cd;e]}{}^{f]}
	& =
	\frac{1}{n-4} \delta^{[a}_{[c} W^{bf]}{}_{de];i}{}^i
	+\frac{2}{n-4} \delta^{[a}_{[c} W^{|i|b}{}_{de];i}{}^{f]}
	\\&
	+\frac{2}{n-4} \delta^{[a}_{[c} W^{bf]}{}_{|i|d;e]}{}^i
	+\frac{4}{(n-3)(n-4)}\delta^{[a}_{[c}\delta^b_d W^{f]i}{}_{e]j;i}{}^j
\end{split}
\end{equation}

This equation can be rewritten in the form $T^{[abf]}{}_{[cde]}=0$ and
it is easy to show that the left hand side is trace free,
i.e. $T^{[abe]}{}_{[cde]}=0$.  It has been shown by Lovelock~\cite{lovelock70}
that for such  trace-free tensors the equation
\begin{equation}\label{5dlovelock}
	T^{[abf]}{}_{[cde]}=0
\end{equation}
is trivially satisfied in five dimensions (and less), i.e. it is a
simple algebraic identity.  It can also be shown by working out the
identity
\begin{equation}
	0=T^{ijk}{}_{[cde}\delta^a_i\delta^b_j\delta^f_{k]}
\end{equation}
which is true in five dimensions (and less) because antisymmetrising
over more than five indices makes the expression vanish; then the
trace-free properties give~\eqref{5dlovelock}.  Equation
\eqref{nocyclicrestrict} is therefore not a constraint in five
dimensions.

To see that it is an effective restriction in six dimensions (and
higher) we can choose a local Cartesian coordinate system and its
associated basis.  We then examine the component corresponding to
$(a,b,f,c,d,e)=(1,2,3,4,5,6)$.  Then equation \eqref{nocyclicrestrict}
becomes
\begin{equation}
	W^{[12}{}_{[45;6]}{}^{3]}
	= 0
\end{equation}
Choosing $W^{12}{}_{45}=W^1{}_4{}^2{}_5 = \sin(x_3)\sin(x_6)$ and all
other components zero except for those needed to give the right
symmetries~\eqref{Wsym} for $W_{abcd}$, then the left hand side
becomes $\pm \cos(x_3)\cos(x_6)$ (the sign depends on the signature of
the metric) which is clearly not zero.

%
%
%

\subsection{Five dimensions (and higher)}

If we now differentiate \eqref{WLd2c1sflat} once again and
antisymmetrise on all free upper indices we obtain
\begin{equation}\label{constraint}
	W^{[ab}{}_{cd;i}{}^{|i|e]}
	+2W^{[ab}{}_{i[c;d]}{}^{|i|e]}
 	+{{4}\over{(n-3)}}
		\delta^{[a}_{[c}W^{b|i}{}_{d]j;i}{}^{j|e]}
	= 0
\end{equation}

It is easy to show that the left hand side is \emph{not} identically
zero in $n>4 $ dimensions. One straightforward way is simply to choose
a local Cartesian coordinate system and its associated basis and
choosing $W^{12}{}_{34}=W^1{}_3{}^2{}_4=\sin(x_5)$ and all other
components zero except for those needed to give the right symmetries.
Then the component on the left hand side corresponding to
$(a,b,c,d,e)=(1,2,3,4,5)$ becomes
\begin{equation}
	W^{12}{}_{34;5}{}^{55}=-\cos(x_5)
\end{equation}
which is not zero.

Another more interesting method is to note that for the subclass of
tensors satisfying $W^a{}_{bcd;a}=0$, the constraint
\eqref{constraint} reduces to
\begin{equation}
	W^{ab}{}_{[cd}{}_{;}{}^i{}_{|i|e]}
	= 0
\end{equation}
which is clearly not identically zero, in $n>4$ dimensions, even when
$W^a{}_{bcd;a}=0$. So, by virtue of \eqref{WL}, there is an effective
restriction \eqref{constraint} imposed on $W_{abcd}$ in dimensions
$n>4$, in flat space.

\subsection{Four dimensions}

In four dimensions, there can be no restrictions on which tensors
$W_{abcd}$ possesses a potential since, as stated in section
\ref{fourdimproof}, in four dimensions any such tensor with the
symmetries \eqref{Wsym} can be given by a Lanczos potential.  However,
it is not immediately obvious from discussion above that the
expressions presented here are not constraints in four dimensions, as
we require.

The constraint \eqref{nocyclicrestrict} is not valid in four
dimensions because of factors of $n-4$ in the denominators.  However,
equation \eqref{WLd2c1sflat} does not contain any $L_{abc}$ and is
valid in four dimensions.  On the other hand \eqref{WLd2c1sflat} can
be rearranged as
\begin{equation}
	  \Bigl(W^{[ab}{}_{[cd}\delta^{i]}_{j]}\Bigr){}^{;j}{}_{i} = 0.
\end{equation}
It can be shown that the left hand side is also identically zero; this
is because the expression in parenthesis
$\Big(W_{[ab}{}^{[cd}\delta^{i]}_{j]}\Bigr)$ belongs to a special
class of dimensionally dependent identities found by
Lovelock~\cite{lovelock70}; these are known to be satisfied \emph{in
four dimensions}.

Hence~\eqref{WLd2c1sflat} is identically satisfied in four dimensions,
and since~\eqref{constraint} is simply a derivative
of~\eqref{WLd2c1sflat} it is also identically zero in four dimensions.

\section{Lanczos potentials for Weyl candidates in\\ curved space}


\subsection{Six dimensions and higher}

It is possible to carry through the same steps that lead to
\eqref{nocyclicrestrict} but in curved space.  Then we get
\eqref{nocyclicrestrict} plus product terms of $L_{abc}$ and curvature
tensors.  By rearranging we are able to use \eqref{WLsimp} to
eliminate some $L_{abc}$, but unfortunately not all; there remain
product terms involving $L_{abc}$ and Weyl curvature tensors:
\begin{equation}\label{curvedbig}
\begin{split}
	W{}^{[ab}&{}_{[cd;e]}{}^{f]}
	 =
	   L_{[cd}{}^{[a;|i|} C^{bf]}{}_{e]i}
	-  L_{[cd}{}^{i;[a} C^{bf]}{}_{e]i}
	-2 L_{[c}{}^{i[a}{}_{;d} C^{bf]}{}_{e]i}
	-2 L_{[c}{}^{i[a;b} C^{f]}{}_{|i|de]}
\\&
	-2 L^{[a|i|}{}_{[c}{}^{;b} C^{f]}{}_{|i|de]}
	+  L_{[c}{}^{i[a} C^{bf]}{}_{de];i}
	+  L^{[a|i|}{}_{[c} C^{bf]}{}_{de];i}
\\&
	-\frac{2}{n-4}\delta^{[a}_{[c}\biggl(
		   L_{de]}{}^{|i;j|} C^{bf]}{}_{ij}
		+2 L_d{}^{|i|b;|j|} C^{f]}{}_{|j|e]i}
		+2 L^{b|i|}{}_d{}^{;|j|} C^{f]}{}_{|i|e]j}
\\&\qquad
		+  L_d{}^{|ij|}{}_{;e]} C^{bf]}{}_{ij}
		+  L^{b|ij|;f]} C_{de]ij}
		+  L_d{}^{|ij|}{}_{;j} C^{bf]}{}_{e]i}
		+  L^{b|ij|}{}_{;|j|} C^{f]}{}_{|i|de]}
\\&\qquad
		+  L_d{}^{|ij|}{}_{;i} C^{bf]}{}_{e]j}
		+  L^{|ij|}{}_d{}^{;b} C^{f]}{}_{e]ij}
		-  L^{|ij|b}{}_{;d} C^{f]}{}_{e]ij}
		+  L^{|ij|b}{}_{;|i|} C^{f]}{}_{|j|de]}
\\&\qquad
		-  L^{|ij|b;f]} C_{de]ij}
		+  L_d{}^{|ij|} C^{bf]}{}_{e]j;i}
		+  L^{b|ij|} C_{de]}{}^{f]}{}_{j;i}
		-\frac{1}{2} L^{|ij|}{}_d C^{bf]}{}_{|ij|;e]}
\\&\qquad
		-\frac{1}{2} L^{|ij|b} C_{de]ij}{}^{;f]}
		+\frac{2}{n-3} L_d{}^{|i|b} C^{f]}{}_{|i|ej;}{}^{j}
		+\frac{2}{n-3} L^{b|i|}{}_d C_{e]i}{}^{f]}{}_{j;}{}^{j}
	\biggr)
\\&
	-\frac{2}{(n-3)(n-4)} \delta^{[a}_{[c}\delta^b_d \biggl(
		 2 L_{e]}{}^{|ij;k|} C^{f]}{}_{jki}
		+2 L^{f]ij;k} C_{|ijk|e]}
		-  L^{|ij|}{}_{e]}{}^{;|k|} C_{ijk}{}^{f]}
\\&\qquad
		-2 L^{|ij|f];k} C_{|ijk|e]}
		+2 L^{|ijk|}{}_{;e]} C_{ijk}{}^{f]}
		+2 L^{|ijk|;f]} C_{|ijk|e]}
		+  L^{|ijk|} C_{|ijk|e]}{}^{;f]}
\\&\qquad
		+  L^{|ijk|} C_{|ijk|}{}^{f]}{}_{;e]}
		+\frac{2}{n-3} L_{e]}{}^{|ij|} C^{f]}{}_{ijk;}{}^{k}
		+\frac{2}{n-3} L^{f]ij} C_{e]ijk;}{}^{k}
\\&\qquad
		+\frac{n-2}{n-3} L^{|ij|}{}_{e]} C_{ij}{}^{f]}{}_{k;}{}^{k}
		+\frac{n-2}{n-3} L^{|ij|f]} C_{|ij|e]k;}{}^{k}
	\biggr)
\\&
	+\frac{4}{(n-3)^2(n-4)} \delta^{[a}_{[c}\delta^b_d\delta^{f]}_{e]}
		L^{ijk} C_{ijkl;}{}^{l}
\\&
	+\frac{1}{(n-2)(n-3)(n-4)} \delta^{[a}_{[c}\delta^b_d\delta^{f]}_{e]}
		C_{ijkl} W^{ijkl}
\\&
	+\frac{1}{n-4} \delta^{[a}_{[c} W^{bf]}{}_{de];i}{}^i
	+\frac{2}{n-4} \delta^{[a}_{[c} W^{|i|b}{}_{de];i}{}^{f]}
	+\frac{2}{n-4} \delta^{[a}_{[c} W^{bf]}{}_{|i|d;e]}{}^i
\\&
	+\frac{4}{(n-3)(n-4)}\delta^{[a}_{[c}\delta^b_d W^{f]i}{}_{e]j;i}{}^j
	-\frac{2}{n-2}W^{[a}{}_{[cd}{}^b \tilde{R}^{f]}{}_{e]}
\\&
	-\frac{4}{(n-2)(n-4)}\delta^{[a}_{[c}W^{b|i|f]}{}_{d}\tilde{R}_{e]i}
	-\frac{4}{(n-2)(n-4)}\delta^{[a}_{[c}W^{b}{}_{de]}{}^{|i|}\tilde{R}^{f]}{}_i
\\&
	-\frac{4}{(n-2)(n-3)(n-4)}\delta^{[a}_{[c}\delta^b_dW^{f]i}{}_{e]}{}^j\tilde{R}_{ij}
\end{split}
\end{equation}
where $\tilde{R}_{ab}$ is the trace free Ricci curvature tensor and $R$
the scalar curvature.

Restricting our considerations to conformally flat manifolds,
i.e. $C_{abcd}=0$, gives
\begin{equation}\label{nocyclicrestrictcurved}
\begin{split}
	W^{[ab}{}_{[cd;e]}{}^{f]}
	& =
	\frac{1}{n-4} \delta^{[a}_{[c} W^{bf]}{}_{de];i}{}^i
	+\frac{2}{n-4} \delta^{[a}_{[c} W^{|i|b}{}_{de];i}{}^{f]}
	\\&
	+\frac{2}{n-4} \delta^{[a}_{[c} W^{bf]}{}_{|i|d;e]}{}^i
	+\frac{4}{(n-3)(n-4)}\delta^{[a}_{[c}\delta^b_d W^{f]i}{}_{e]j;i}{}^j
	\\&
	-\frac{2}{n-2}W^{[a}{}_{[cd}{}^b \tilde{R}^{f]}{}_{e]}
	-\frac{4}{(n-2)(n-4)}\delta^{[a}_{[c}W^{b|i|f]}{}_{d}\tilde{R}_{e]i}
	\\&
	-\frac{4}{(n-2)(n-4)}\delta^{[a}_{[c}W^{b}{}_{de]}{}^{|i|}\tilde{R}^{f]}{}_i
	\\&
	-\frac{4}{(n-2)(n-3)(n-4)}\delta^{[a}_{[c}\delta^b_dW^{f]i}{}_{e]}{}^j\tilde{R}_{ij}.
\end{split}
\end{equation}
where we note that all the terms containing $L_{abc}$ explicitly have
disappeared.

As in the flat case, this expression can be rearranged in the form
$T^{[abf]}{}_{[cde]}=0$ where $T^{[abf]}{}_{[cde]}$ is trace free, so
this is not a restriction in five dimensions.  When specialised to
flat space this restriction is the same as~\eqref{nocyclicrestrict} so
we know that this is an effective restriction in six dimensions and
higher even though we do not know if this is an effective restriction
for \emph{all} conformally flat manifolds.

\subsection{Five dimensions (and higher)}

Carrying through the same steps that led to \eqref{constraint} but in
a general curved space gives an expression that is unmanageable in
size.  However, if we assume constant curvature, i.e. $C_{abcd}=0$ and
$\tilde{R}_{ab}=0$, we do get a manageable restriction.  With those
assumptions we get
\begin{equation}\label{constraintcurved}
\begin{split}
	W^{[ab}{}_{cd;i}{}^{|i|e]}
	+2W^{[ab}{}_{i[c;d]}{}^{|i|e]}
 	+\frac{4}{n-3} \delta^{[a}_{[c}W^{b|i}{}_{d]j;i}{}^{j|e]}
	&\\
	-\frac{2(n-4)}{n(n-1)}RW^{[a}{}_c{}^b{}_d{}^{;e]}
	+\frac{4}{n(n-1)}R\delta^{[a}_{[c}W^{b|i|e]}{}_{d];i}
	&= 0
\end{split}
\end{equation}
where $R$ is the scalar curvature and we note that all the terms
containing $L_{abc}$ explicitly have disappeared.

When specialised to flat space this restriction is the same
as~\eqref{constraint} and we can therefore conclude that this is an
effective restriction in five dimensions and higher.  However, we do
not know whether this is an effective restriction for \emph{all}
constant curvature spaces.

The restrictions \eqref{nocyclicrestrictcurved} and
\eqref{constraintcurved} can be somewhat rearranged if the
symmetry~\eqref{Wcycl} is used but the effect is only cosmetic, see
comment~\ref{note:parallell} in next section.


\section{Summary and discussion} \label{sec:notes}

What has been proved here can be summarised in the following theorems.

\begin{theorem}
Let $M$ be a metric $n$-dimensional differentiable flat manifold,
$n\geq 6$, with any signature.  Then tensors $W_{abcd}$ --- with the
properties \eqref{Wsym} --- which fail to satisfy
\eqref{nocyclicrestrict} cannot be given via~\eqref{WL} by a
three times differentiable potential $L_{abc}$ with the properties
\eqref{Lsym}.
\end{theorem}

\begin{theorem}
Let $M$ be a metric $n$-dimensional differentiable flat manifold,
$n\geq 5$, with any signature.  Then tensors $W_{abcd}$ --- with the
properties \eqref{Wsym} --- which fail to satisfy
\eqref{constraint} cannot be given via~\eqref{WL} by a four
times differentiable potential $L_{abc}$ with the properties
\eqref{Lsym}.
\end{theorem}

\begin{theorem}
Let $M$ be a metric $n$-dimensional differentiable manifold, $n\geq
6$, with zero Weyl curvature and any signature.  Then tensors
$W_{abcd}$ --- with the properties \eqref{Wsym} --- which fail to
satisfy \eqref{nocyclicrestrictcurved} cannot be given via~\eqref{WL}
by a three times differentiable potential $L_{abc}$ with the
properties \eqref{Lsym}.
\end{theorem}

\begin{theorem}
Let $M$ be a metric $n$-dimensional differentiable manifold, $n\geq
5$, with constant curvature and any signature.  Then tensors
$W_{abcd}$ --- with the properties \eqref{Wsym} --- which fail to
satisfy \eqref{constraintcurved} cannot be given via~\eqref{WL} by a
four times differentiable potential $L_{abc}$ with the properties
\eqref{Lsym}.
\end{theorem}

In the above calculations the properties \eqref{Lcycl} and
\eqref{Wcycl} have not been used.  Therefore we can also conclude that
in these theorems we can relax the condition on $W_{abcd}$
to~(\ref{Wsym}a,b,d) and on $L_{abc}$ to~\eqref{Lasym}.  This more
general result is of importance when we compare our results with the
results of Bampi and Caviglia~\cite{bampi83} for the parallel problem
discussed in Comment~\ref{note:parallell} below.

%
%

\vspace{10pt}
A number of  points should be emphasised:

\note
One must, of course, pose the question whether there are other
possible generalisations of \eqref{WLorig} which would be more
successful than \eqref{WL}.  Making the assumptions that the potential
should be a three index tensor and having the symmetries \eqref{Lsym},
then already the very simple and most obvious choice
\begin{equation}\label{WLguess}
	W_{abcd} = 4L_{abc;d}
\end{equation}
leads to the same generalisation via the following steps.  The
antisymmetry of the last two indices gives
\begin{equation}
	W_{abcd} = W_{ab[cd]} = 4L_{ab[c;d]}
\end{equation}
The symmetry \eqref{Wpsym} gives
\begin{equation}
	W_{abcd}
	=
	\frac{1}{2} W_{abcd} + \frac{1}{2} W_{cdab}
	=
	2L_{ab[c;d]} +2L_{cd[a;b]}
\end{equation}
Subtracting off the trace of this equation gives us \eqref{WL}.

It is difficult to imagine another generalisation which cannot be
reduced to~\eqref{WL}.

\note
The most obvious way to see that there exist some $W_{abcd}$ which
have a potential is to choose a tensor $L_{abc}$ and then \emph{define}
a tensor $W_{abcd}$ via~\eqref{WL}.  Obviously the $W_{abcd}$ so
obtained has this Lanczos potential $L_{abc}$.

Of course this result does not rule out the possibility of \emph{some}
significant subclasses of these tensors $W_{abcd} $ having such a
potential --- even in flat spaces.  For instance, suppose that a
tensor $B_{abcd}$, with the properties \eqref{Wasym} and
\eqref{Wcycl}, also satisfies the Bianchi-like equation
$B_{ab[cd;e]}=0 $; in flat space, in all dimensions, it is well known
that such tensors $B_{abcd}$ \emph{with this additional property} can
be written as $B_{abcd} = 2h_{a[c;d]b}- 2h_{b[c;d]a}$.  Consider next
the case where $B_{abcd}$ is a Weyl candidate i.e., $B_{abcd}\equiv
W_{abcd}$, and it follows that $W_{abcd}$, satisfying \eqref{Wsym} and
the additional property $W_{ab[cd;e]}=0 $, can always be written in
the form \eqref{WL} with the appropriate 3-potential
$L_{abc}=h_{c[a;b]}$.  Clearly for $W_{abcd}$ with such properties,
the constraints \eqref{nocyclicrestrict} and \eqref{constraint} is
trivially satisfied, and so they are no longer effective constraints
in this case.

\note
We have been considering Weyl candidates in general, and have shown
explicitly that the set of Weyl candidates for which a Lanczos
potential can be defined in dimensions $n$, where $n>4$, does not
exhaust the complete set; since the Weyl curvature tensor is itself a
special case from the set of all Weyl candidates we are unable to draw
any direct conclusions for \emph{Weyl curvature tensors} from our
results above.  So the question whether \emph{all Weyl curvature
tensors} have a Lanczos potential is still open.

\note
We have only shown directly that Lanczos potentials for \emph{general}
Weyl candidates fail to exist in flat space ($n\geq 5$).  In
conformally flat space ($n\geq 5$) and constant curvature ($n\geq 6$)
we have obtained restrictions on which Weyl candidates may have a
Lanczos potential even though we have not proved that these
restrictions are effective for all spaces in the respective class.
This leaves the possibility open that there are some \emph{special
subclasses} of these classes where all Weyl candidates have a Lanczos
potential.

\note
The existence proofs of the Lanczos potential in $n>4$ dimensions
would need to be established differently than the existence proofs for
the four dimensional case since they have to exclude some tensors
$W_{abcd}$ or alternatively exclude some spaces.

Rather than seeing this as an existence problem maybe one should see
this as a problem of characterising the class of tensors $W_{abcd}$
that can be defined through~\eqref{WL} for $n>4$ dimensions.

\note \label{note:parallell}
In a \emph{parallel problem}, regarding the existence of a potential
$\tilde L_{abc}$ with the property \eqref{Lasym}, for a 4-tensor
$\tilde W_{abcd}$ with the properties \eqref{Wasym}, \eqref{Wpsym} and
\eqref{Wtrace}, Bampi and Caviglia~\cite{bampi83} have claimed that
such a potential $\tilde L_{abc}$ generally does not exist in spaces
of $n>6$, but does exist for spaces of dimension $n=4,5,6$.  However,
it is important to note that although this parallel problem is
equivalent to the original Weyl candidate problem when $n=4$, it is
\emph{not} equivalent to the original problem when $n\ne4$, and so
this parallel result \emph{cannot} be transferred directly to the
original problem for a Weyl candidate (or Weyl tensor) when
$n\ne4$. Therefore the conclusion --- that a Lanczos potential
$L_{abc}$ for the Weyl tensor $C_{abcd}$ exists only in spaces of
dimension 4,5,6 (see Roberts~\cite{roberts89},
Edgar~\cite{edgar94a,edgar94b}) --- which has been drawn from the
result of this parallel problem is unjustified.

In fact in the calculations of Bampi and Caviglia~\cite{bampi83}
leading to the parallel result just mentioned, there seems to be a
simple computational error in the very last step of their argument;
when this is corrected, their result for the parallel problem should
be that a potential does exist generally for spaces of dimension
$n=4,5$, but not for spaces where $n>5$.

The parallel problem considered by Bampi and Caviglia~\cite{bampi83}
was only a way of getting the results for the original problem.  In
this paper, as well as in the paper of Bampi and Caviglia, the
original problem is the primary target.  However, we have deliberately
avoided using the extra properties \eqref{Wcycl} and \eqref{Lcycl} in
the original problem compared to the parallel problem; \emph{therefore
our results are valid for the parallel problem as well}.

So our results for the parallel problem agree with Bampi and Caviglia
for $n\geq 6$; but since we also obtain effective restrictions on
$W_{abcd}$ in the parallel problem for $n=5$ we seem to contradict
the results of Bampi and Caviglia.

Since Bampi and Caviglia only discuss the cases for $n>4$ in passing
in their conclusion and also stress that their results in this case
are of a generic nature, more work needs to be done to determine
precisely their result and its correspondence to ours in 5 dimensions.

\note \label{note:nosurp}
The non-existence of the Lanczos potential, in dimensions $n>4$,
should not really come as a surprise.  Firstly, the existence proofs
of Bampi and Caviglia~\cite{bampi83}, Illge~\cite{illge88} and
Andersson and Edgar~\cite{andersson00} depend explicitly and crucially
on dimension four; it has proved impossible to write these proofs in a
manner which could be used in arbitrary dimension.  Secondly, it has
been shown by Illge~\cite{illge88}, via spinors, and by
Edgar~\cite{edgar94a}, via tensors for arbitrary signature in 4
dimensions, that substitution of (1) into the Bianchi equations gives
the very simple differential equation
\begin{equation}\label{Lwavesimp}
	\nabla^2 L_{abc} = 0
\end{equation}
for zero Ricci tensor and Lanczos gauges.  On the other hand, for
dimensions $n>4$ substitution of \eqref{WL} into the Bianchi equations
gives a much more complicated differential equation,
\begin{multline}\label{Lwave}
	\nabla^2 L_{abc}
	+\frac{2(n-4)}{n-2}L_{[a}{}^{d}{}_{|c;d|b]}\\
	-2L_{[b}{}^{ed}C_{a]dec}
	+\frac{1}{2}C_{deab}L^{de}{}_{c}
	-\frac{4}{n-2}g_{c[a}C_{b]fed}L^{fed}
	= 0
\end{multline}
for zero Ricci tensor and Lanczos gauges (Edgar and
H\"oglund~\cite{edgar97}).  Equation \eqref{Lwavesimp} links up with
classical existence and uniqueness results for second order partial
differential equations; the more complicated equation \eqref{Lwave} is
not suitable --- because of its second term --- for applying these
classical results.  Thirdly those algorithms in the literature which
enable particular Lanczos potentials to be calculated explicitly are
constructed in a manner in which the dimension four is crucial (often
using spinors); it seems impossible to generalise these algorithms to
higher dimensions.


%
\end{document}